\documentclass[twocolumn]{aastex62}
\usepackage{graphicx,url}
\usepackage{amsmath,amssymb}
\usepackage{hyperref}
\hypersetup{colorlinks,
	linkcolor=blue, % Color for normal internal links, eg: blue
	urlcolor=magenta, % Color for linked URLs, eg: magenta
	anchorcolor=blue,
	citecolor=blue
}
\usepackage{bm}
\usepackage{xcolor}
\usepackage{mathrsfs}
\usepackage{threeparttable}

\begin{document} 

\title{\LARGE\bfseries Parkes transient events: I. Database of single pulses, initial results and missing FRBs}

\author{S.-B. Zhang}
\affil{Purple Mountain Observatory, Chinese Academy of Sciences, Nanjing 210008, China}
\affil{University of Chinese Academy of Sciences, Beijing 100049, China}
\affil{CSIRO Astronomy and Space Science, PO Box 76, Epping, NSW 1710, Australia}
\affil{International Centre for Radio Astronomy Research, University of Western Australia, Crawley, WA 6009, Australia}

\author{G. Hobbs}
\affil{CSIRO Astronomy and Space Science, PO Box 76, Epping, NSW 1710, Australia}

\author{C. J. Russell}
\affil{CSIRO Scientific Computing Services, Australian Technology Park, Locked Bag 9013, Alexandria, NSW 1435, Australia}

\author{L. Toomey}
\affil{CSIRO Astronomy and Space Science, PO Box 76, Epping, NSW 1710, Australia}

\author{S. Dai}
\affil{CSIRO Astronomy and Space Science, PO Box 76, Epping, NSW 1710, Australia}

\author{J. Dempsey}
\affil{CSIRO Information Management \&\ Technology, Box 225, Dickson, ACT 2602, Australia}

\author{R. N. Manchester}
\affil{CSIRO Astronomy and Space Science, PO Box 76, Epping, NSW 1710, Australia}

\author{S. Johnston}
\affil{CSIRO Astronomy and Space Science, PO Box 76, Epping, NSW 1710, Australia}

\author{L. Staveley-Smith}
\affil{International Centre for Radio Astronomy Research, University of Western Australia, Crawley, WA 6009, Australia}
\affil{ARC Centre of Excellence for All Sky Astrophysics in 3 Dimensions (ASTRO 3D)}

\author{X.-F. Wu}
\affil{Purple Mountain Observatory, Chinese Academy of Sciences, Nanjing 210008, China}

\author{D. Li}
\affil{University of Chinese Academy of Sciences, Beijing 100049, China}
\affil{National Astronomical Observatories, Chinese Academy of Sciences, A20 Datun Road, Chaoyang District, Beijing 100101, China}
\affil{FAST Collaboration, CAS Key Laboratory of FAST, NAOC, Chinese Academy of Sciences, Beijing 100101, China}

\author{Y.-Y. Yang}
\affil{FAST Collaboration, CAS Key Laboratory of FAST, NAOC, Chinese Academy of Sciences, Beijing 100101, China}
\affil{School of Physics and Electronic Science, Guizhou Education University, Guiyang 550018, China}

\author{S.-Q. Wang}
\affil{University of Chinese Academy of Sciences, Beijing 100049, China}
\affil{CSIRO Astronomy and Space Science, PO Box 76, Epping, NSW 1710, Australia}
\affil{Xinjiang Astronomical Observatory, Chinese Academy of Sciences, Urumqi, Xinjiang 830011, China}

\author{H. Qiu}
\affil{CSIRO Astronomy and Space Science, PO Box 76, Epping, NSW 1710, Australia}
\affil{Sydney Institute for Astronomy, School of Physics, University of Sydney, Sydney, NSW 2006, Australia}

\author{R. Luo}
\affil{CSIRO Astronomy and Space Science, PO Box 76, Epping, NSW 1710, Australia}

\author{C. Wang}
\affil{CSIRO Data61, Sydney, NSW 2015, Australia}

\author{C. Zhang}
\affil{University of Chinese Academy of Sciences, Beijing 100049, China}
\affil{CSIRO Astronomy and Space Science, PO Box 76, Epping, NSW 1710, Australia}
\affil{National Astronomical Observatories, Chinese Academy of Sciences, A20 Datun Road, Chaoyang District, Beijing 100101, China}
\affil{CSIRO Data61, Sydney, NSW 2015, Australia}

\author{L. Zhang}
\affil{University of Chinese Academy of Sciences, Beijing 100049, China}
\affil{CSIRO Astronomy and Space Science, PO Box 76, Epping, NSW 1710, Australia}
\affil{National Astronomical Observatories, Chinese Academy of Sciences, A20 Datun Road, Chaoyang District, Beijing 100101, China}

\author{R. Mandow}
\affil{CSIRO Astronomy and Space Science, PO Box 76, Epping, NSW 1710, Australia}
\affil{Centre for Astrophysics and Supercomputing, Swinburne University of Technology, Mail H30, PO Box 218, Hawthorn, VIC 3122, Australia}

\begin{abstract}
A large number of observations from the Parkes 64\,m-diameter radio telescope, recorded with high time resolution, are publicly available. We have re-processed all of the observations obtained during the first four years (from 1997 to 2001) of the Parkes Multibeam receiver system in order to identify transient events and have built a database that records the 568,736,756 pulse candidates generated during this search. We have discovered a new fast radio burst (FRB), FRB~010305, with a dispersion measure of 350$\pm$5\,\,cm$^{-3}\,$pc and explored why so few FRBs have been discovered in data prior to 2001. After accounting for the dispersion smearing across the channel bandwidth and the sky regions surveyed, the number of FRBs is found to be consistent with model predictions.  We also present five single pulse candidates from unknown sources, but with Galactic dispersion measures. We extract a diverse range of sources from the database, which can be used, for example, as a training set of data for new software being developed to search for FRBs in the presence of radio frequency interference.
\end{abstract}

\section{Introduction}
\label{sec:intro}

%Parkes introduction
The Parkes 64 m-diameter radio telescope has been used to discover more than half of the known pulsars through surveys such as those described by \citet{Manchester01} and \citet{Hobbs04}, as well as the first Rotating Radio Transients~\citep[RRATs;][]{McLaughlin06} and Fast Radio Bursts~\citep[FRBs;][]{Lorimer07,Thornton13}. With updated receivers and signal processing equipment such as the Multibeam (MB)~\citep{Staveley-Smith96} and the Ultra Wideband Low (UWL) ~\citep{Hobbs19} receivers and their corresponding backend instrumentation, the telescope systems have remained versatile and continue to make new discoveries.   

The majority of the Parkes high-time resolution data sets are publicly available from CSIRO's data archive\footnote{CSIRO Data Access Portal, DAP, \url{https://data.csiro.au}}~\citep{Hobbs11}. This archive allows new algorithms to be tested and then applied to the large data volumes~\citep[recent examples of new discoveries found in the archive include][]{Pan16,Zhang18,Zhang19}. The archive contains more than 100 observing projects, with each observing semester for each project stored as a data collection. Approximately 600 such data collections are now available for public access.  This archive provides a very long data set ($\sim$29 years) with stable observing systems.

It is likely that astronomical sources remain to be found in the archive.  We have therefore started a project in which we will search all the archival data in a self-consistent manner for transient signals and create a database of all the single pulses detected. We have chosen to start by processing the search mode 
data sets available in the data archive that were recorded using the Multibeam receiver between 1997 and 2001.  The year 2001 was chosen as the initial cut-off date as this enables us to study the apparent lack of FRB discoveries before this date\footnote{The data on the DAP are accessed through specific data collections, which include all observations for a specific project during one observing semester. The last data set used in this paper corresponds to the 2001 May observing semester.}. 

%FRBs introduction
FRBs are bright, single pulses of millisecond duration and expected to have extragalactic origin.
Almost 100 FRBs have now been published\footnote{\url{http://www.frbcat.org}}~\citep{Petroff16}. The Canadian Hydrogen Intensity Mapping Experiment (CHIME) has the highest rate of new FRB discoveries~\citep{CHIME19a,CHIME19b} and the Australian Square Kilometre Array Pathfinder (ASKAP) has demonstrated the ability to localise the sources within their host galaxies~\citep{Bannister19}. However, telescopes with high sensitivity such as Parkes continue to make FRB discoveries~\citep[e.g.,][]{Oslowski19} and extend the distribution of FRBs on occurred time, dispersion measure (DM) and expected luminosity etc..

%FRB gaps
\begin{figure*}
\begin{center}
\begin{tabular}{l}
\includegraphics[width=15cm,angle=0]{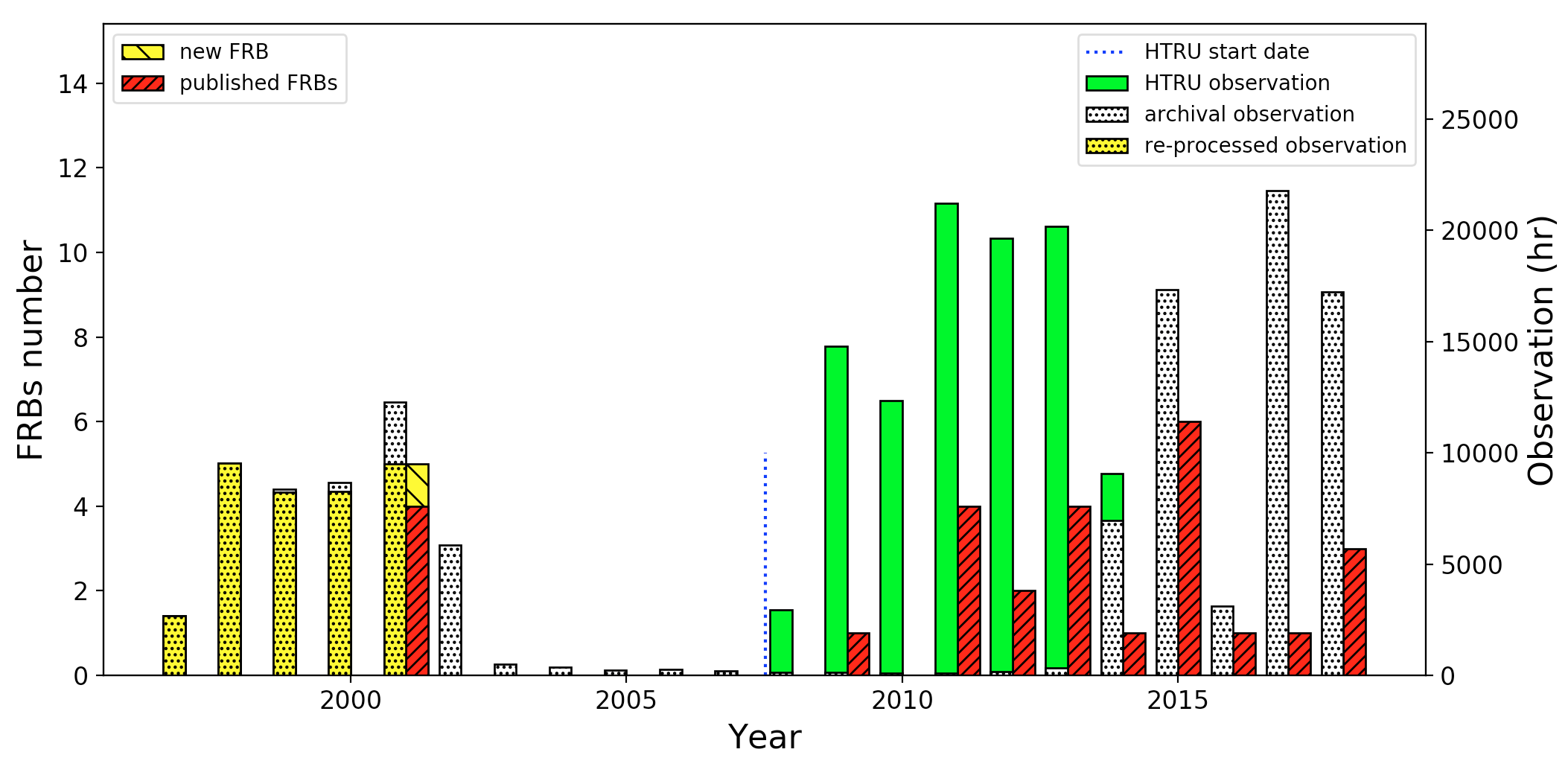} 
\end{tabular}
\caption{The FRBs detected by the Parkes telescope (shown in red and yellow hashed bars) and the number of observations contained in the data archive (dotted bars), our database (yellow, dotted bars) and discovered in the HTRU survey (lime bars). The vertical dotted line indicates the time that the HTRU survey commenced. The HTRU survey data sets are not currently available on the data archive and therefore not included in the dotted bars. The number of HTRU observations was estimated by the archive of Parkes observing schedules    
(\protect\url{https://www.parkes.atnf.csiro.au/observing/schedules/prev_schedules.php}) and the allocated time (i.e., 7092\,hr) for the HTRU survey. Note that this value is $\sim$ 39\% larger than the expected survey length of $\sim$5100\,hr~\citep{Keith10}. }
\label{figure:Parkes_FRB}
\end{center}
\end{figure*}

In Figure~\ref{figure:Parkes_FRB} we show the event times for FRB detections with Parkes overlaid on the epochs of the use of the Multibeam Receiver.  From 1997 to 2010, only five FRBs were discovered, and four of them occurred in the first half of 2001~\citep{Burke-Spolaor14,Zhang19,Keane12,Lorimer07} and one in 2009~\citep{Champion16}.  
From Figure~\ref{figure:Parkes_FRB} it is clear that there are two intervals (from 1997 to 2001 and from 2002 to 2009) in the FRB detections where no FRB have been detected. 

Between 2001 and 2009, the Parkes telescope was involved in large scale pulsar timing experiments and spectral-line and continuum surveys. The major pulsar survey during this time was the~\citet{Bates11} survey that made use of a multibeam methanol receiver. We only have a significant number of 20\,cm multibeam observations for the first apparent gap in the number of detected FRBs, i.e., from 1997 to mid-2001.  
The data archive contains 38,190\,hr of on-sky integration time across all beams using the Multibeam Receiver during this period; see yellow dotted bars in Figure~\ref{figure:Parkes_FRB}.  
We would expect events to be evenly distributed throughout the span of the data, but note that these early observations had lower frequency resolution than more recent Parkes observations.  This limits the maximum dispersion measure (DM) and minimum width for detectable FRBs; the implications of this are described in Section~\ref{sec:why_missing}.  The choice of DM ranges and trials in previous analyses, methods of radio frequency interference (RFI) mitigation and signal to noise ratio (S/N) estimation could also have led to FRBs being missed~\citep{Keane19}.

%database introduction
%Our database contains 568,736,756 events. Clearly the majority of these events are not FRBs and the database can be used to study single pulses from pulsars and RRATs (the results of such a study will be presented in a subsequent paper), to search for repeating events from known FRBs, and to monitor transient radio frequency interference. The database can also be used to produce training data sets for machine learning algorithms.  Of course, some of the events are of unknown origin.  
%Here we describe five single pulse candidates that we cannot easily explain. 

In this paper we describe the details of the observations and data reduction in Section~\ref{sec:data} before explaining the structure of the database in Section~\ref{sec:db}.  The properties of the new discoveries are presented in Section~\ref{sec:result} and discussed in Section~\ref{sec:dis}. We conclude in Section~\ref{sec:conclu}.  The database and associated software are separately available for public download from CSIRO as described in Appendix~\ref{sec:usage}.  Appendix~\ref{sec:usage} also includes  instructions for use of the database.

\section{Observation and Data Reduction}
\label{sec:data}
%\subsection{Observations}
\label{sec:obs}
%data sets summary
The data sets used in the database were all obtained with the primary goal of discovering pulsars.  
%data sets parameters 
The channelised and polarization-summed signals were one-bit sampled and recorded using an Analogue Filter Bank (AFB) system. Each project semester has been saved as a data collection, named with the project code identifier and the semester period. 
%each project
The observing projects that have contributed observations include: P268, the Galactic Plane Parkes Multibeam Survey~\citep[e.g.,][]{Manchester01,Hobbs04,Lorimer06}; P269, a deep survey of the Large and Small Magellanic Clouds~\citep{Crawford01,Manchester06}; P309, a survey of intermediate Galactic latitudes~\citep{Edwards01} and P360 and P366 which were high-latitude pulsar surveys~\citep{Burgay06,Jacoby09}. All the observations used the  20\,cm Multibeam Receiver on the Parkes 64\,m-diameter radio telescope from May 1997 to August 2001 and together led to the discovery of more than eight hundred new pulsars. There are 21 collections in our current analysis. The central frequency, bandwidth and number of channels are 1374\,MHz, 288\,MHz and 96 respectively. We list the sampling times and integration time of these observations in Table~\ref{table:collections}. 

\begin{table*}
\caption{Data collections used in this study, the number of data files, on-sky integration time and the sample time of each collection.  Here we also list the number of detections of known pulsars and FRBs during our processing. The DOIs of the collections are listed in Appendix~\ref{raw_data_dois}. Note that, for an unknown reason, some data collections contained observations outside of the observing semester specified. We include all the available data here and in our processing.}
\begin{center}
\begin{tabular}{lrrccc}
\hline
\hline
Collection       &  Number    &  Integration  & $t_{\rm samp}$   & Detected & Detected   \\
  Name           &  of Files   &  Time(h)      &  ($\mu$s)        & Pulsars      & FRBs         \\
\hline
P268$-$1997AUGT  &     4,615  &  2,668    &   250  &  35  &    \\
P268$-$1998JANT  &     4,446  &  2,594    &   250  &  58  &    \\
P268$-$1998MAYT  &     5,525  &  3,128    &   250  &  71  &    \\
P268$-$1998SEPT  &     3,926  &  2,287    &   250  &  49  &    \\
P309$-$1998SEPT  &    20,959  &  1,514    &   125  &  57  &    \\
P268$-$1999JANT  &     1,638  &    956    &   250  &  19  &    \\
P309$-$1999JANT  &    23,940  &  1,714    &   125  &  49  &    \\
P268$-$1999MAYT  &     3,198  &  1,866    &   250  &  34  &    \\
P309$-$1999MAYT  &    25,803  &  1,875    &   125  &  48  &     \\
P268$-$1999SEPT  &     3,120  &  1,813    &   250  &  34  &     \\
P268$-$2000JANT  &     2,405  &  1,403    &   250  &  19  &     \\
P268$-$2000MAYT  &     3,068  &  1,782    &   250  &  27  &     \\
P269$-$2000MAYT  &     2,082  &  2,444    &  1000  &   1  &     \\
P268$-$2000OCTT  &     1,859  &  1,084    &   250  &  18  &      \\
P269$-$2000OCTT  &       951  &  1,276    &  1000  &   2  &      \\
P268$-$2001JANT  &     1,131  &    660    &   250  &  20  &      \\
P269$-$2001JANT  &     2,258  &  2,520    &  1000  &   7  &  FRB~010724; FRB~010312 \\
P360$-$2001JANT  &    17,718  &  1,277    &   125  &  15  &  FRB~010125             \\
P268$-$2001MAYT  &     1,820  &  1,062    &   250  &  13  &       FRB~010621             \\
P360$-$2001MAYT  &    13,885  &  1,018    &   125  &  10  &      \\
P366$-$2001MAYT  &    44,122  &  3,249    &   125  &  19  &   FRB~010305             \\
\hline
Total            &   188,469   & 38,190   &        &  385  &   5              \\
\hline
\hline
\end{tabular}
\end{center}
\label{table:collections}
\end{table*}

%\subsection{Single Pulse Search}
%\label{sec:sing}
%RFI mitigation
The data were processed using the pulsar searching software package \emph{\sc presto}\footnote{\url{http://www.cv.nrao.edu/~sransom/presto/}. Note that an update to \emph{\sc PRESTO} was required in order to handle the archived 1-bit data files from Parkes correctly.}~\citep{Ransom01} on CSIRO's high performance computer facilities. Strong narrow-band and short-duration broadband radio frequency interference (RFI) were identified and marked using the \emph{\sc presto} routine \emph{\sc rfifind}. We used a 1\,s integration time for our RFI identification and the default cut-off to reject time-domain and frequency-domain interference in our pipeline\footnote{Some bright signals, e.g., the pulses of Vela pulsar whose flux density at 1400\,MHz is 1050\,mJy~\citep{Jankowski18}, are sometimes also marked as RFI by this pipeline.}. We also recorded the single pulse candidates at zero DM (without any RFI mitigation) to enable long-term RFI studies with our database. 

%de-disperse 
The data sets were dedispersed at DM values that were determined by the \emph{\sc ddplan.py} algorithm with the option \emph{\sc r}  (``acceptable time resolution'') set to be 0.5\,ms and based on the central frequency, bandwidth, number of channels and sampling time. The DM range was from $1$ to $5000\,\,$cm$^{-3}\,$pc. Data were then de-dispersed at each of the trial DMs using the \emph{\sc prepdata} routine with RFI removal based on the mask file produced by \emph{\sc rfifind}.  
To avoid missing bright burst events, we used the option \emph{\sc noclip} which disabled auto-clipping of data during all the processing steps and the option \emph{\sc b} for the \emph{\sc single\_pulse\_search.py} routine which disabled the check for bad-blocks. 
This is an extremely parallel computational challenge so a Perl script, \emph{\sc nproc}, was used to accelerate the de-dispersion and single pulse searching steps by distributing individual computing jobs across multiple CPUs.

%single pulse search
Single pulse candidates with S/N larger than seven were identified using the \emph{\sc single\_pulse\_search.py} routine for each de-dispersed time series and for different boxcar filtering {parameters} with filter widths of 1, 2, 3, 4, 6, 9, 14, 20, 30, 45, 70, 100, 150, 220 and 300 samples. We use the definition of $\sigma$ as presented by \textsc{presto} as our S/N value. A signal with S/N above seven is unlikely to have been generated by receiver noise.  
%and also usually presents clear structure in the un-dedispersed data (see Section~\ref{sec:why_missing} for more details). 

%{\bf All candidates with S/N above seven were stored in the database; as a further step we searched for interesting astrophysical signals with a S/N threshold of ten that is universally applied in FRBs search~\citep{Champion16}. }
%The S/N of a candidate can {\bf be increased if a} a finer grid of DM trials {\bf is used}. As a supplement to the prior results {\bf we reanalyzed the data corresponding to candidates with S/N values between nine and ten using a finer grid. We used DM values in a range $\pm$5\,\,cm$^{-3}\,$pc centered at the candidate's DM and with a DM step of 0.1\,\,cm$^{-3}$pc.} The database has not recorded the results from this supplementary search.

\section{Database of the Single Pulses}
\label{sec:db}   
\begin{figure*}
\begin{center}
\begin{tabular}{l}
%\fbox{\includegraphics[width=17cm,angle=0,trim={91mm 53mm 84mm 55mm},clip]{spc-db.pdf}}
\includegraphics[width=17cm,angle=0,trim={91mm 53mm 84mm 55mm},clip]{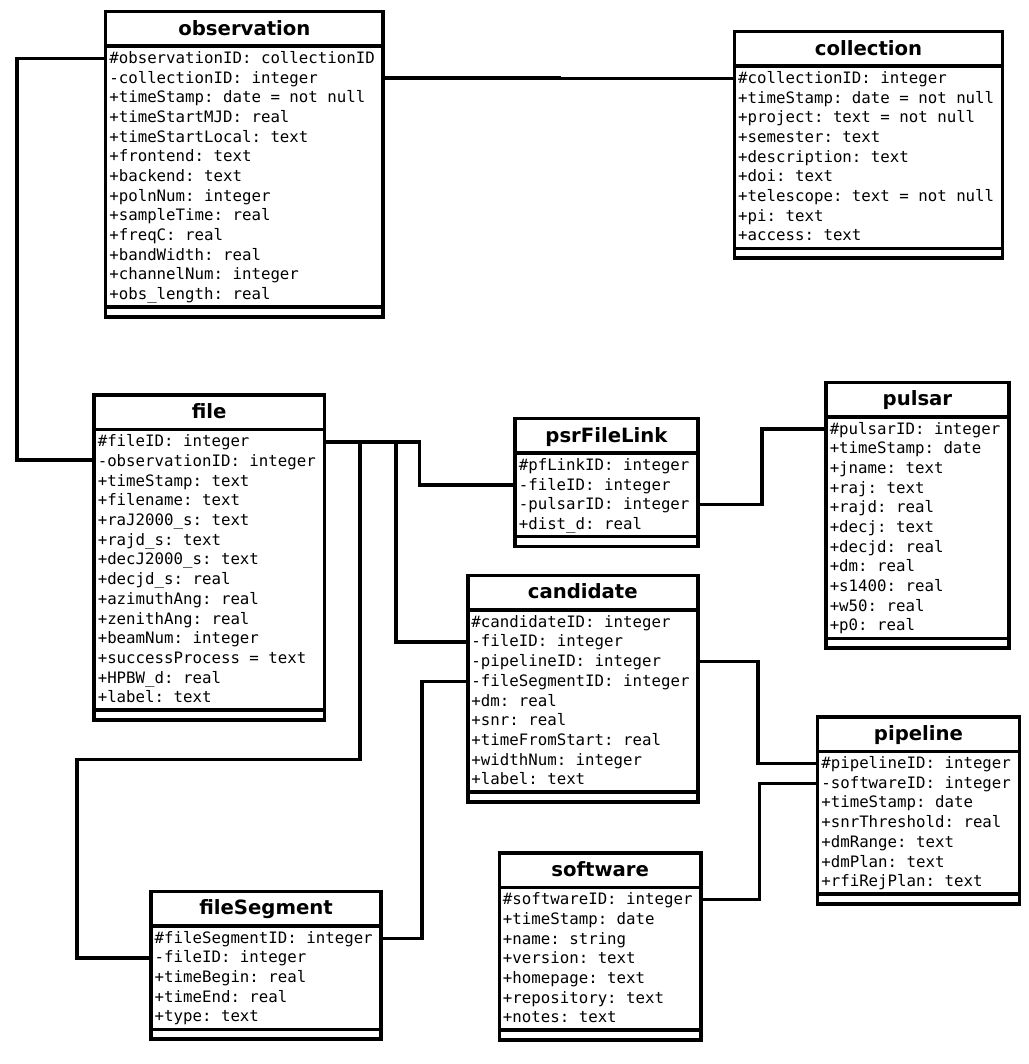}
\end{tabular}
\caption{Chart of the database schema} 
\label{figure:Database_schema}
\end{center}
\end{figure*}

We have created an SQLite\footnote{\url{https://www.sqlite.org/}} database to store the transient events.   The database contains nine tables that store the data and three indices, which are used to speed up searches of the database. The database schema and description of the parameters are listed in Table~\ref{table:schema} and graphically displayed in Figure~\ref{figure:Database_schema}.

%introduction of each table
In brief, each data collection is identified by the \textsc{collection} table. This contains information such as the project identifier and the observing semester.  The data collections contain multiple observations.  The \textsc{observation} table is therefore linked to the corresponding collection and contains the start time of the observation and details of the receiver and signal processor system used to acquire the data as well as information such as the sampling time, central observing frequency and the bandwidth. Most of the observations currently in the database were acquired with all 13 beams of the multibeam receiver. A data file is produced for each beam of the receiver for each observation (in a minority of cases, the observer chose only to record information from the central beam of the multibeam receiver). Information on these data files are stored in the \textsc{file} table which includes the beam number and beam pointing direction.

We have provided details (such as software versions) of the software and pipeline used to extract the single pulse events in the \textsc{software} and \textsc{pipeline} tables (note that, in this paper, we present results from the two pipelines described above, one that searches for astrophysical signals using a DM range from $1$ to $5000\,\,$cm$^{-3}\,$pc, and the other that searches for radio frequency interference at a DM value of 0 \,\,cm$^{-3}\,$pc, but in the future we will compare the results from different processing pipelines using different software packages). The single pulse candidates are saved in the table \textsc{candidate}, which contains information such as the time, DM, S/N and the pulse width corresponding to each candidate.

The huge volume of candidates implies that it would be impossible to inspect them all by eye. Each candidate has the arrival time at the highest observing frequency (the start time of the event) and at the lowest frequency (the end time). To aid in inspecting these candidates, we have grouped different candidates showing common features. For instance, candidates with adjacent DMs and overlapping start and end times, often derive from the same wide-profile signal. We have therefore grouped all the candidates that lie within the start and end time of a given event.
The table \textsc{fileSegment} records the fileID and the start and end time for these groupings\footnote{The database only contains the information for the candidates, but we plan to also provide the actual segments of raw data in a later version. Some examples are shown in the additional support directory of \texttt{db\_fileSeg} mentioned in Section\ref{sec:stat}.}.

We also provide a table (\textsc{pulsar}) of known pulsars based on the latest (i.e., version 1.62) ATNF Pulsar Catalogue\footnote{\url{http://www.atnf.csiro.au/research/pulsar/psrcat}}~\citep{Manchester05}. Where possible (i.e., where the beam position is correct and single pulses have been detected) we have linked specific observations to known pulsars using the \textsc{psrFileLink} table in the database.

\begin{table*}
\footnotesize
\caption{The parameters stored in the database.}
\begin{center}
\begin{tabular}{lll}
\hline
\hline
Name          &   Type          &   Description         \\
\hline
{\bf collection} &             &   Unit of the data sets, normally contains the observations of one project semester         \\
collectionID     &   integer   &   The primary key         \\
timeStamp        &   date      &   Time stamp of insertion         \\
project          &   text      &   Observing project code identifier          \\
semester         &   text      &   Semester for the observing project          \\
description      &   text      &   Primary goal of the observation         \\
doi              &   text      &   Digital object identifier linked to the data sets         \\
telescope        &   text      &   Telescope used to take the observation          \\
pi               &   text      &   Principal Investigator of the observing project         \\
access           &   text      &   data access status: available or embargo         \\
\hline
{\bf observation} &             &   One complete observing tracking          \\
observationID     &   integer   &   The primary key         \\
collectionID      &   integer   &   Identifier for the collection, link to the table of collection         \\
timeStamp         &   date      &   Time stamp of insertion\\           
timeStartMJD      &   real      &   Start MJD of the observation in UTC time         \\
%timeStartLocal    &   text      &   Local time of the observation \\
frontend          &   text      &   Receiver used to acquire the data         \\
backend           &   text      &   Signal processor used for observation         \\
polnNum           &   integer   &   Number of polarizations  \\
sampleTime        &   real      &   Sample interval for SEARCH-mode data (s)         \\
freqC             &   real      &   Centre frequency (MHz)          \\
bandWidth         &   real      &   Observation bandwidth (MHz)         \\
channelNum        &   integer   &   Number of frequency channels         \\
obs\_length       &   real      &   The full duration of the observation (s)         \\
\hline
{\bf file}     &             &   A file contains the observation from one beam         \\
fileID         &   integer   &   The primary key          \\
observationID  &   integer   &   Identifier for the observation, link to the table of observation         \\
timeStamp      &   date      &   Time stamp of insertion         \\
filename       &   text      &   Name of the file         \\
raJ2000\_s     &   text      &   Right ascension in J2000 coordinates of the pointing centre of the beam (hh:mm:ss.ssss)          \\
rajd\_s        &   real      &   Right ascension in J2000 coordinates of the pointing centre of the beam (deg)          \\
decJ2000\_s    &   text      &   Declination in J2000 coordinates of the pointing centre of the beam  (dd:mm:ss.sss)          \\
decjd\_s       &   real      &   Declination in J2000 coordinates of the pointing centre of the beam (deg)         \\
azimuthAng     &   real      &   Azimuth angle (deg)          \\
zenithAng      &   real      &   Zenith angle (deg)          \\
beamNum        &   integer   &   Beam number for multibeam systems (1=central beam)         \\
successProcess &   text      &   Here will note the details if the file process failed         \\
HPBW\_d        &   real      &   Half power beamwidth of the beam (deg)         \\
label          &   text      &   Special file will be labelled         \\
\hline
{\bf software} &             &   Software used to obtain the candidates            \\
softwareID     &   integer   &   The primary key          \\
timeStamp      &   date      &   Time stamp of insertion          \\
name           &   text      &   name of the software         \\
version        &   text      &   version of the software         \\
homepage       &   text      &   homepage of the software         \\
repository     &   text      &   URL to software/code repository         \\
notes          &   text      &   Special changes when using the software\tnote{g}          \\
\hline
{\bf pipeline} &             &   Pipeline built to process the data and obtain candidates.          \\
pipelineID     &   integer   &   The primary key          \\
softwareID     &   integer   &   Identifier for the software, link to the table of software         \\
timeStamp      &   date      &   Time stamp of insertion         \\
snrThreshold   &   real      &   S/N threshold to get the candidates         \\
dmRange        &   text      &   Searched DM range         \\
dmPlan         &   text      &   Method to decide the DM trials       \\
rfiRejPlan     &   text      &   Method to reject the RFI           \\
\hline
\hline
\end{tabular}
\end{center}
\label{table:schema}
\end{table*}

\begin{table*}
{\footnotesize
\begin{centering}
{\small  {\bf Table~\ref{table:schema}.} \emph{continued}\\ }
\begin{tabular}{lll}
\hline
\hline
Name          &   Type          &   Description         \\
\hline
{\bf candidate} &             &   The details of the candidates obtained during our search        \\
candidateID     &   integer   &   The primary key          \\
fileID          &   integer   &   Identifier for the searched file, link to the table of file         \\
pipelineID      &   integer   &   Identifier for the used pipeline, link to the table of pipeline         \\
fileSegmentID   &   integer   &   Identifier for the file segment contains the candidate, link to the table of fileSegment         \\
dm              &   real      &   DM where the candidate detected (cm$^{-3}\,$pc)         \\
snr             &   real      &   S/N of the candidate         \\
timeFromStart   &   real      &   time when the candidate detected from the start of the observation, here actually recorded     \\

                &             &   time of arrival of the signal at the highest frequency. \\
widthNum        &   integer   &   number of samples for the boxcar filtering to get the candidate         \\
label           &   text      &   the candidates of special event will be labelled         \\
\hline
{\bf fileSegment} &            &   A file segment contains the grouped candidates in one group         \\
fileSegmentID     &  integer   &   The primary key          \\
fileID            &  integer   &   Identifier for the searched file, link to the table of file         \\
timeBegin         &  real      &   time of the segment begin         \\
timeEnd           &  real      &   time of the segment end          \\
type              &  text      &   type of the events in the segment          \\
\hline
{\bf pulsar} &            &   pulsars parameters provided by the latest ATNF pulsar catalogue         \\
pulsarID     &  integer   &   The primary key         \\
timeStamp    &  date      &   Time stamp of insertion         \\
jname        &  text      &   Pulsar name based on J2000 coordinates         \\
raj          &  text      &   Right ascension (J2000) (hh:mm:ss.s)          \\
rajd         &  real      &   Right ascension (J2000) (deg)          \\
decj         &  text      &   Declination (J2000) (dd:mm:ss)          \\
decjd        &  real      &   Declination (J2000) (deg)         \\
dm           &  real      &   Dispersion measure (cm$^{-3}\,$pc)         \\
s1400        &  real      &   Mean flux density at 1400 MHz (mJy)         \\
w50          &  real      &   Width of pulse at 50\% of peak (ms)         \\
p0           &  real      &   Barycentric period of the pulsar (s)          \\
\hline
{\bf psrFileLink} &             &   Files containing pulsars with single-pulses with a S/N $\ge 8$ will be linked to the related pulsar          \\
pfLinkID          &   integer   &   The primary key         \\
fileID            &   integer   &   Identifier for the file         \\
pulsarID          &   integer   &   Identifier for the related pulsar         \\
dist\_d           &   real      &   Distance between the location the pulsar and the pointing centre of the beam (deg)         \\
\hline
\hline
\end{tabular}
\end{centering}
}
\end{table*}

\section{Results}
\label{sec:result}
 
We initially investigated all the files that contained candidates with S/N$>$8 in the DM-time plane which were plotted by the usage \texttt{plot\_TimeDM} described in Section~\ref{ap_usage}. Any candidate that was seen as a isolated burst and has DM larger than DM$_{\rm MW}$ was selected for further analysis. Some files were significantly affected by RFI and therefore it was impossible to view every candidate with this S/N threshold. We therefore also inspected the groupings of candidates. For these groups we only viewed the candidate with the highest S/N present within that group. 
The S/N of these candidates were slightly increased using a finer grid of DM trials than in the normal processing. We used DM values in a range $\pm$5\,\,cm$^{-3}\,$pc centered at the candidate's DM and with a DM step of 0.1\,\,cm$^{-3}$pc. Eventually we obtained the DMs with the highest S/N.
%{\bf For all the candidates presented in this section, we have optimized the DMs to obtain the highest S/N.}

We re-detected all four of the published FRBs. As listed in Table~\ref{table:collections}, they are FRB~010125 at a DM of $786.5\,\,$cm$^{-3}\,$pc in beam 5 with S/N of 17.9, FRB~010312 at a DM of $1163\,\,$cm$^{-3}\,$pc in beam 7 with S/N of 11.0, FRB~010621 at a DM of $749\,\,$cm$^{-3}\,$pc in beam 10 with S/N of 15.8, and FRB~010724 at a DM of $373\,\,$cm$^{-3}\,$pc in three beams (beams 6, 7, and 13) with S/N of 32.0, 15.0 and 24.1.

\begin{table}
\caption{The properties of FRB~010305. The S/N was calculated after removing the RFI based on the mask file produced by \emph{\sc rfifind}. The width was obtained by fitting the integrated pulse profile at its 50\% power point.}
\begin{center}
\begin{threeparttable}
\begin{tabular}{ll}
\hline
\hline
{\bf Observed Properties} \\
Event data UTC                                      &  2001 March 5         \\
Event time UTC, $\nu_{1.374 \rm {GHz}}$             &  12:29:16.02          \\
Event time Local (AEDT), $\nu_{1.374 \rm {GHz}}$    &  23:29:16.02          \\
Pointing R.A. (J2000)                               &  04:57:19.5           \\
Pointing Dec. (J2000)                               &  $-$52:36:24.668      \\
Galactic longitude                                  &  260.06$^{\circ}$     \\
Galactic latitude                                   &  $-$38.34$^{\circ}$   \\
Beam 3 full-width, half-maximum                     &  14.1$'$              \\
DM (cm$^{-3}\,$pc)                                  &  350$\pm$5            \\
Observed width (ms)                                 &  9$\pm$1.5         \\
S/N                                                 &  10.2                 \\
\hline
{\bf Inferred Properties}\\
Peak flux density (Jy)                  &  0.42            \\
Fluence (Jy ms)                         &  3.78            \\
DM$_{\rm MW, YMW16}$ (cm$^{-3}\,$pc)    &  36              \\
Redshift$_{\rm YMW16}$, z               &  0.3\tnote{a}    \\
Distance$_{\rm YMW16}$ (Gpc)            &  1.2\tnote{a}    \\
\hline
\end{tabular}
   \begin{tablenotes}
        \footnotesize
        \item[a]The DM of host galaxy was assumed to be 100\,\,cm$^{-3}\,$pc and the calculation used the YMW16 model~\citep{Yao17}. 
     \end{tablenotes}
\end{threeparttable}
\end{center}
\label{table:properties}
\end{table}

\begin{figure}
\begin{center}
\begin{tabular}{l}
\includegraphics[width=8cm,angle=0]{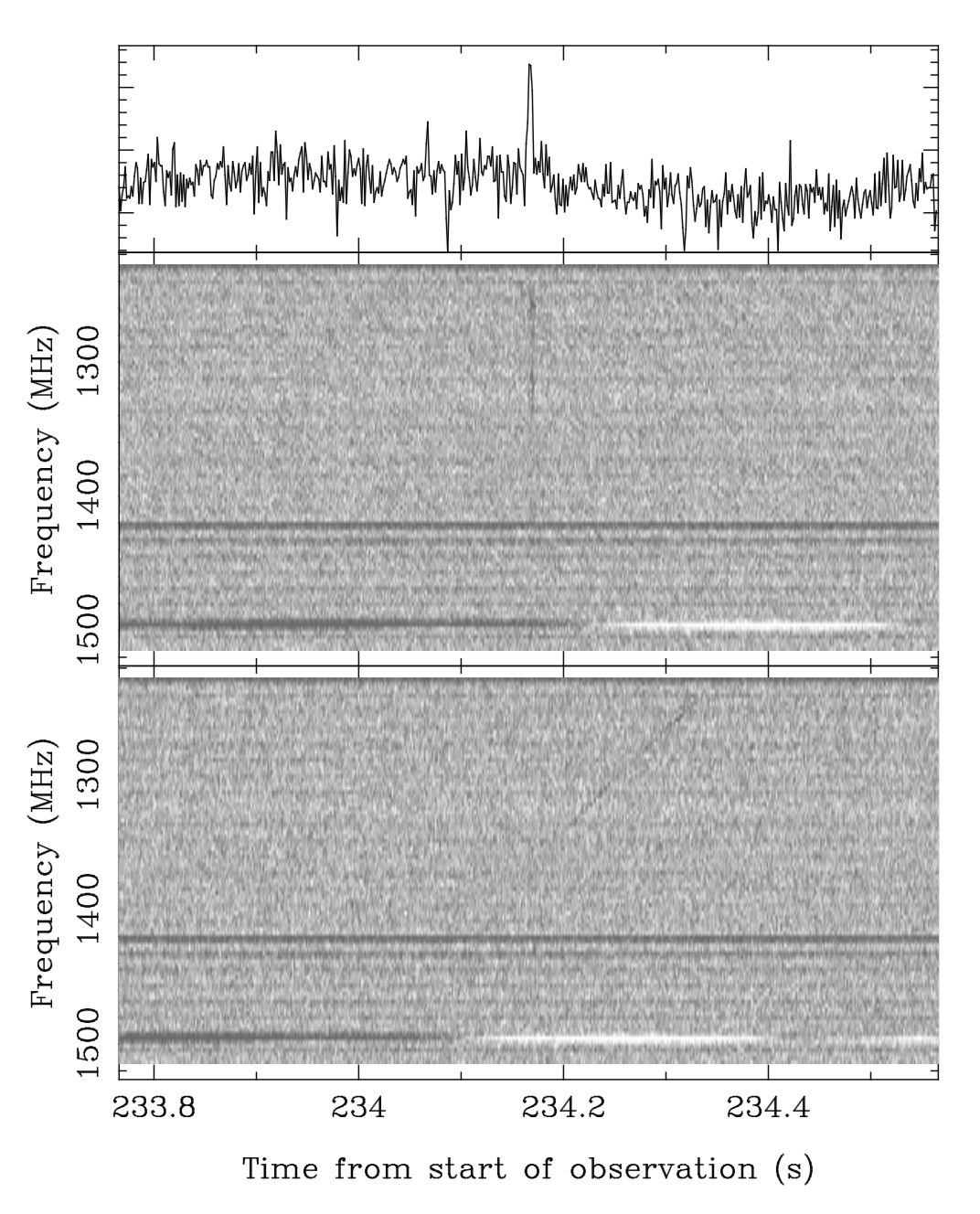} 
\end{tabular}
\caption{Frequency-time plane without de-dispersion (bottom), after de-dispersing with the optimized DM of 350\,cm$^{-3}$pc (central) and integrated pulse profile (top) for FRB~010305.  The time axis shows the time since the start of the observation (March 5, 2001 12:25:22 UTC) with a resolution of 1.5\,ms.} 
\label{figure:FRB010305}
\end{center}
\end{figure}

A new FRB (which, following tradition, we label as FRB~010305) with a DM of 350$\pm$5\,\,cm$^{-3}\,$pc and S/N$=$10.2 was  detected in our search. Figure~\ref{figure:FRB010305} shows the burst in the frequency-time plane and its integrated pulse profile after being de-dispersed at the optimal DM value. The signal is much stronger in the lower part of the observing band, which is similar to FRBs such as FRB~010312~\citep{Zhang19}, FRB~110214~\citep{Petroff19} and FRB~171019~\citep{Shannon18}. Two strong narrow-band RFI signatures in the higher band are clear from the Figure, but they do not affect our confidence in the FRB detection. We are confident that this is a real FRB as our database contains no other unexpected event with a DM value larger than the Galactic DM contribution (with a S/N value between eight and ten).

We list the properties of FRB~010305 in Table~\ref{table:properties}. This FRB, the second-earliest known, was only detected in a single beam (Beam 3) of the Multibeam Receiver, and no clear pulse candidate or RFI occurred around the time of the burst in the remaining 12 beams. Detection in a single beam cannot provide a precise position or fluence.   The coordinates listed are simply the pointing position of the beam. The burst has a width of 9$\pm$2\,ms at its 50\% power point. In the lower half of Table~\ref{table:properties}, we provide various inferred  properties of FRB~010305. The peak flux density was obtained from the single pulse radiometer equation~\citep{Cordes03} and the S/N measurement. The YMW16 electron density model~\citep{Yao17}, which assumes $H_0=67.3\,{\rm km}\,{\rm s}^{-1}\,{\rm Mpc}^{-1}$~\citep{Planck14} and the local intergalactic medium baryon density $n_{\rm IGM} = 0.16\,{\rm m}^{-3}$~\citep{Katz16}, indicates a cosmic distance of 1.2\,Gpc with the assumption of a host galaxy DM of 100\,\,cm$^{-3}\,$pc. In order to search for the possibility of the FRB repeating~\citep{Spitler16}, we identified 3.24\,h of observations whose beam pointing positions were within 1.0\,deg of the position of the beam in which FRB~010305 was detected. No convincing candidates were identified in these other observations.

Our search for new FRBs led to the detection of numerous single, dispersed pulse events with DM values likely to put the source within our Galaxy. Events that are detected in all the multibeam beams simultaneously are commonly referred to as ``perytons'' and they have been identified as being generated when a microwave oven door is opened prematurely~\citep{Petroff15}. We have detected 22 peryton events in total. Their details can be obtained by the tools presented in Section~\ref{sec:db_use_training} and Appendix~\ref{ap_usage}.

Most of the candidates in the database are from locally generated interference or from single pulses of known pulsars. We have manually identified 1,084 observation files that contain 385 unique pulsars with single pulses of S/N $\ge 8$ based on the positions, DMs and estimated periods; the mean observation time for each of these pulsars is $\sim$ 55\,min.  
A detailed analysis of these pulses and RFI detections will be presented elsewhere. Examples are provided in Section~\ref{sec:db_use_training}. 

\begin{figure*}
\begin{center}
\begin{tabular}{ccc}
\includegraphics[width=5.7cm,angle=0]{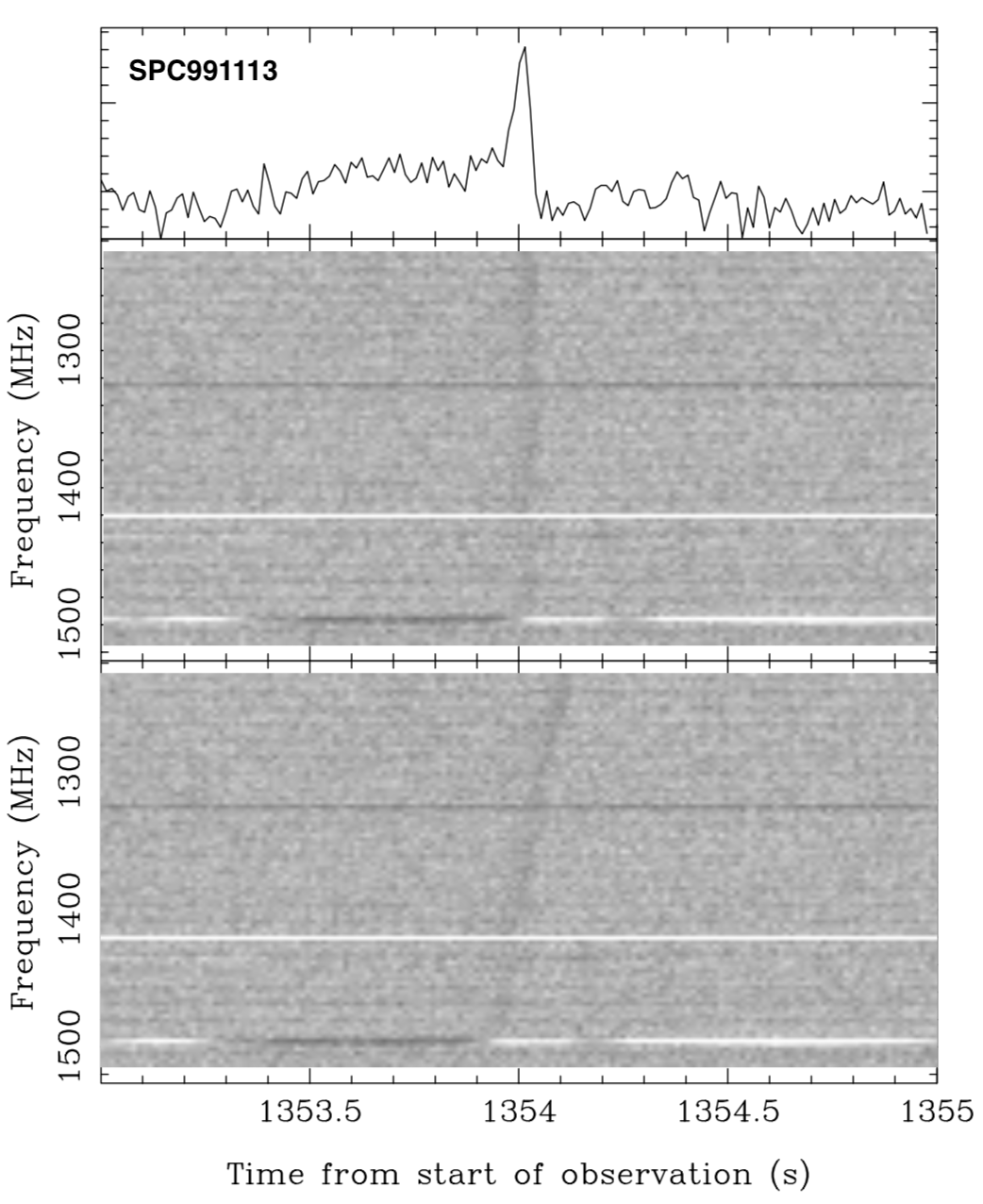} &
\includegraphics[width=5.5cm,angle=0]{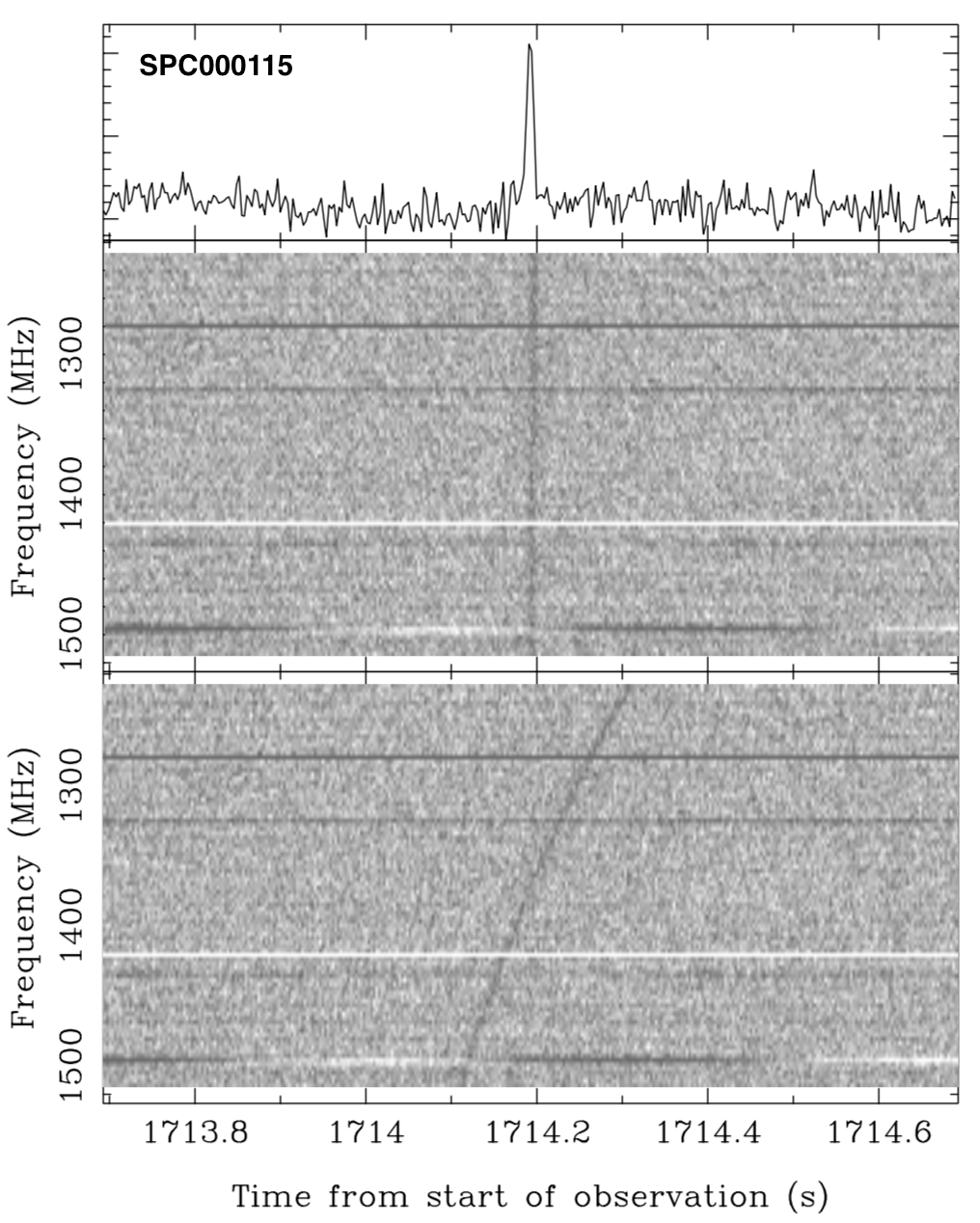} &
\includegraphics[width=5.5cm,angle=0]{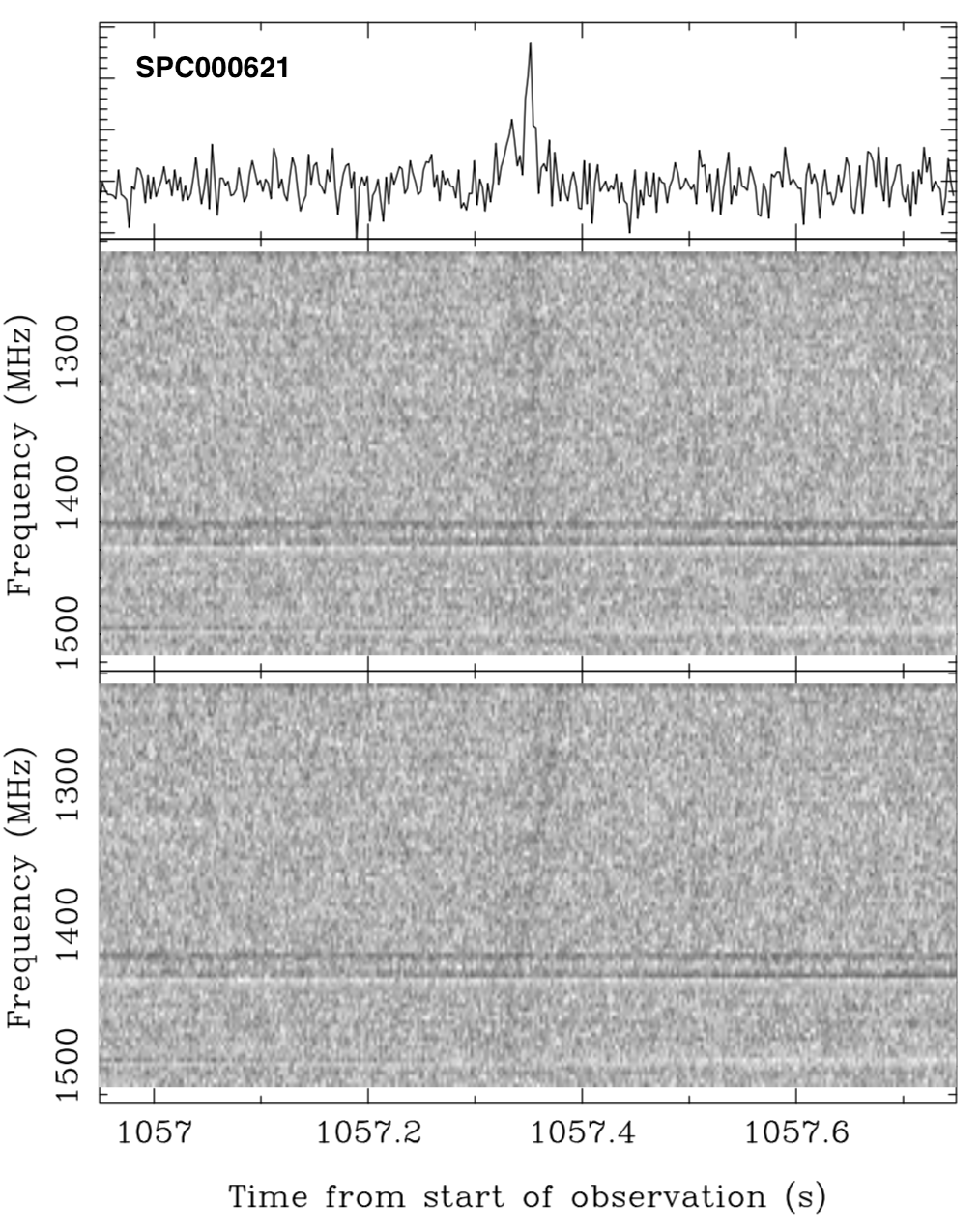} \\
(a) &
(b) &
(c) \\
\end{tabular}
\begin{tabular}{cc}
\includegraphics[width=5.5cm,angle=0]{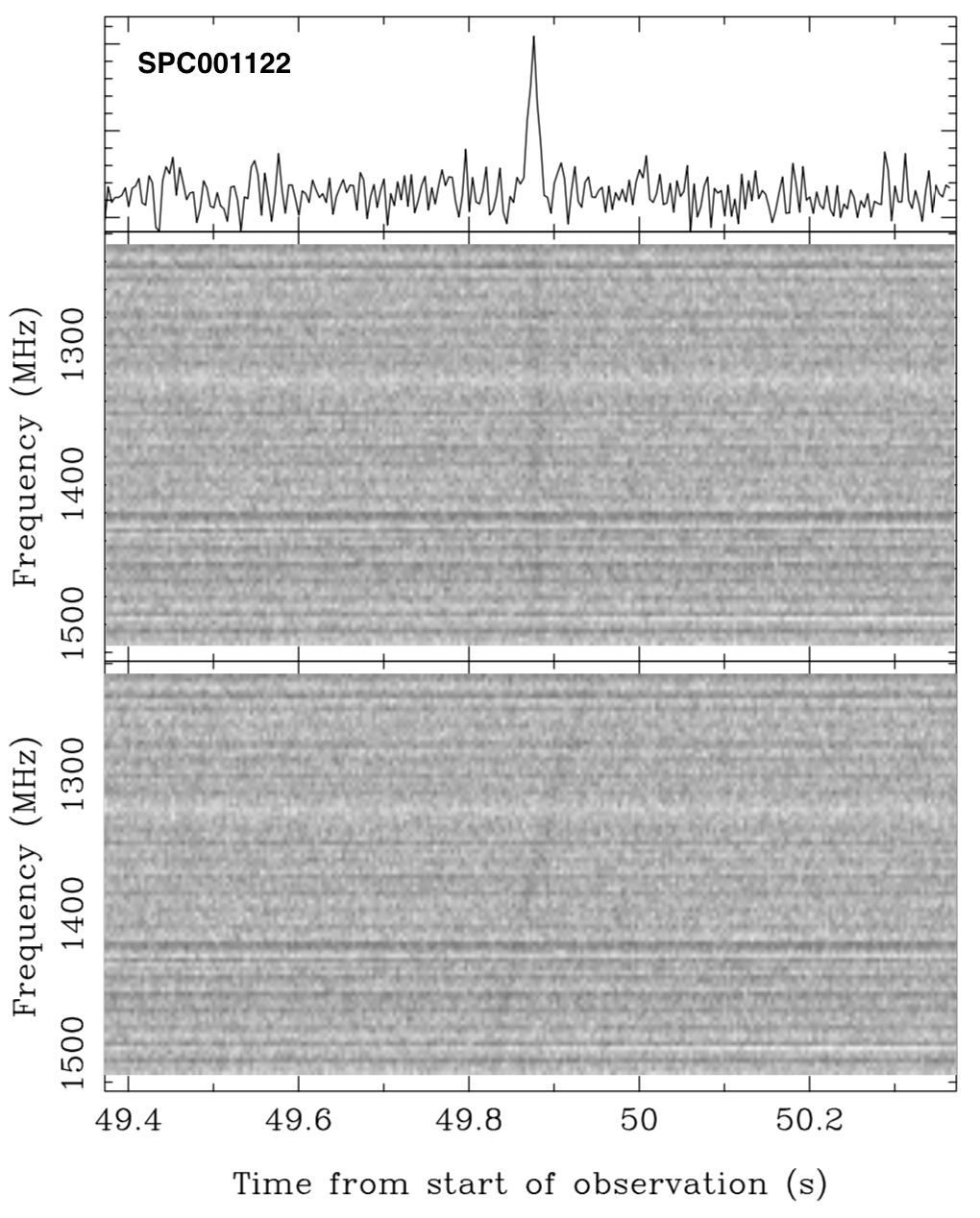} &
\includegraphics[width=5.5cm,angle=0]{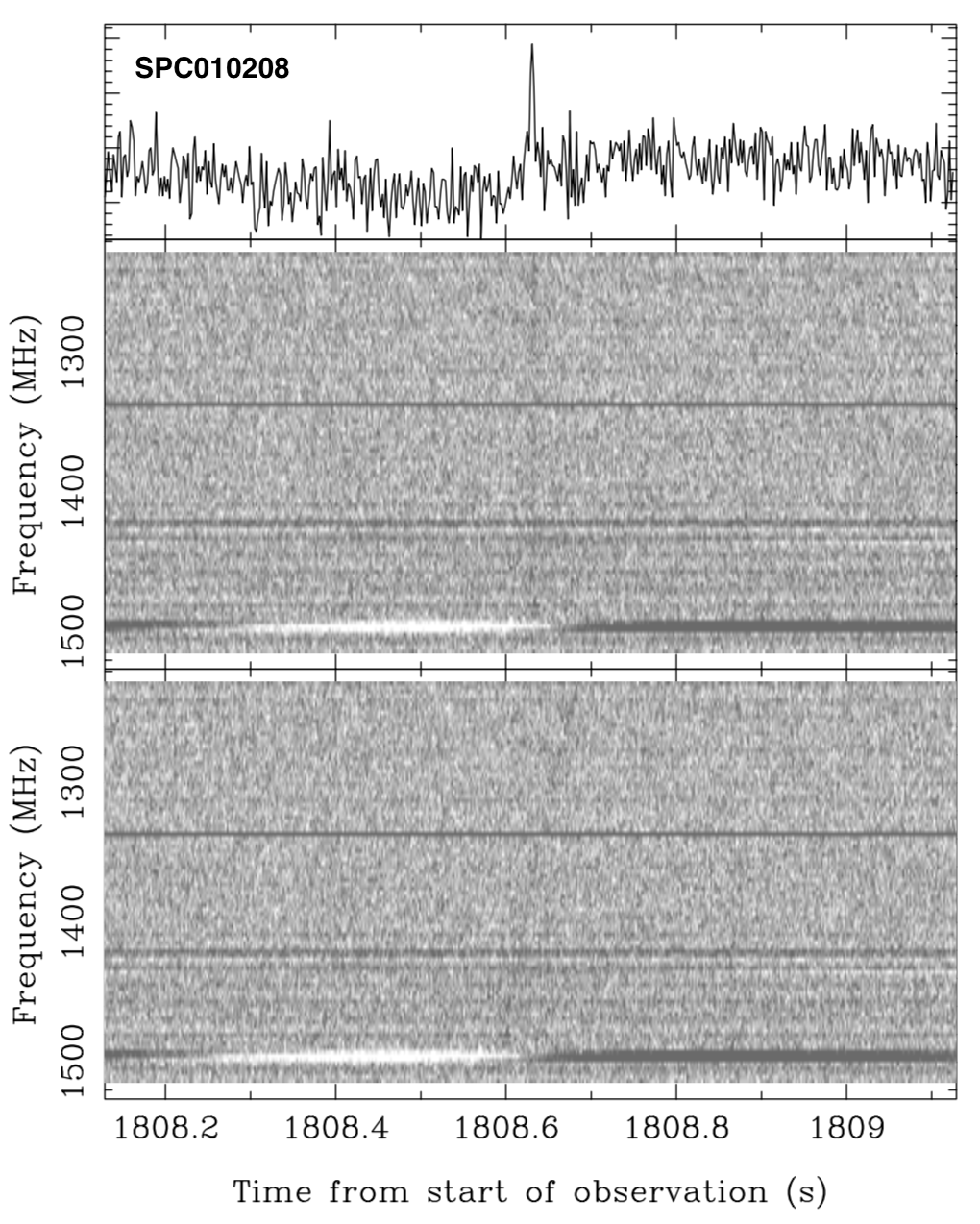} \\
(d) &
(e) \\
\end{tabular}
\caption{Pulse profiles for five single pulse events with Galactic DM values.  The time resolution of (a) SPC~991113, (b) SPC~000115, (c) SPC~000621, (d) SPC~001122, and (e) SPC~010208 are 52, 12, 10, 4, 8\,ms, respectively. For each sub-figure, the bottom  panel shows the frequency-time plane, the central panel shows the event after being de-dispersed at the optimized DMs, whilst the top panel is the integrated pulse profile using an arbitrary flux scale.}
\label{figure:SPC101}
\end{center}
\end{figure*}

Some dispersed events have only been detected in a single beam, have DMs indicating a Galactic source, and the beam pointing directions are not in the direction of known radio pulsars.  Five such events, those with S/N values above 10, are shown in Figure~\ref{figure:SPC101}.  The candidate label includes SPC (denoting ``single pulse candidate'') and the detection date. We list the best-fitting DM, S/N, detecting beam, the beam pointing position in right ascension and declination (and converted to Galactic coordinates), observed width at 50\% power point, inferred peak flux density, fluence, DM$_{\rm MW, YMW16}$ and Distance$_{\rm YMW16}$ of these SPCs in  Table~\ref{table:SPC_properties}.    We note that SPC~991113 may be a bright pulse from the RRAT~J1739$-$2521 with DM of 186.4\,cm$^{-3}\,$pc as their positions are close and the uncertainty for the DM of this SPC is relative large ($\sim$ 26\,cm$^{-3}\,$pc).   
However, the single pulses from that RRAT are not expected to be so bright, nor so broad~\citep{Cui17}. No convincing candidate was obtained from the periodicity search of the files containing these candidates. 

These single pulse candidates have only been detected in a single beam of the receiver suggesting that they are unlikely to be from terrestrial signals.  There are no known pulsars or RRATs in the beam pointing directions and therefore are likely to be new RRATs, or giant pulses from currently unknown pulsars. No more convincing candidates were identified in the same sky directions of them from the data archive.  
We will re-observe these sky regions with the Parkes telescope, but note that all these sky regions could also be observed with the more sensitive MeerKAT~\citep{Jonas16} telescope and, SPC~000621 or SPC~010208 with the FAST telescope~\citep{Jiang19}. However, we note that the positions of these candidates are only poorly determined to date.

\begin{table*}
\caption{The properties of five single pulse events with Galactic DM values. The width was obtained by fitting the integrated pulse profile at its 50\% power point.}
\begin{center}
\begin{tabular}{lccccc}
\hline
\hline
\multicolumn{1}{c}{Entity}                  &  SPC~991113       &  SPC~000115      &  SPC~000621     &  SPC~001122     &  SPC~010208      \\
\hline
Event date UTC                              &  1999 Nov. 13    &  2000 Jan. 15   &  2000 June 21  &  2000 Nov. 22  &  2001 Feb. 08   \\
Event time UTC                              &  04:12:20.92     &  16:28:15.11    &  15:53:14.32   &  20:35:50.82   &  23:17:50.59    \\
Event time Local (AEDT)                     &  15:12:20.92     &  03:28:15.11    &  01: 53:14.32  &  07:35:50.82   &  09:17:50.59    \\
Pointing R.A. (J2000)                       &  17:39:49.6      &  13:28:55.8     &  16:05:35.7    &  08:08:10.6    &  19:05:49.0     \\
Pointing Dec. (J2000)                       &  $-$25:13:16.2   &  $-$58:54:05.9  &  $-$45:45:05.2 &  $-$32:18:11.0 &  $-$01:26:42.1  \\
Galactic longitude ($^{\circ}$)             &  2.48            &  307.77         &  334.59        &  249.96        &  33.28          \\
Galactic latitude ($^{\circ}$)              &  3.05            &  3.62           &  4.85          &  0.20          &  $-$3.86        \\
Detected beam number                        &  12              &  12             &  3             &  7             &  13             \\
DM (cm$^{-3}\,$pc)                          &  203             &  213            &  65.1          &  135.5         &  102.6          \\
DM$_{\rm MW, YMW16}$ (cm$^{-3}\,$pc)        &  407.7           &  357.1          &  294.4         &  549.5         &  330.9          \\
Distance$_{\rm YMW16}$ (kpc)  &  4.8             &  6.1           &  1.9    &  0.44         &   3.9  \\ 
Observed width (ms)                         &  150             &  30             &  30            &  12.5          &  20             \\
S/N                                         &  12.4            &  19.4           &  11.4          &  11.6          &  10.2           \\
Peak flux density (Jy)                      &  0.08            &  0.28           &  0.17          &  0.26          &  0.18           \\
\hline
\end{tabular}
\end{center}
\label{table:SPC_properties}
\end{table*}

%a bit more a bout RFI

\section{Discussion}
\label{sec:dis}

%\subsection{Where are the missing FRBs?}
\label{sec:why_missing}
%why missing FRBs: gl, gb VS observaton length analysis; 
Our primary goal was to process the first four years of Multibeam data from the Parkes telescope in order to understand why very few FRBs were discovered during that time. Between 1997 and June 2001 we now know of five FRBs (including the new event presented in this paper). However, they all (including our new event) occurred in the first half of 2001.  

The expected detection rate for a given set of observations depends upon (1) the DM range searched and (2) the channel bandwidth of the observations.
The data sets between 1997 and mid 2001 contain four independent observing projects.  These observing projects have different Galactic sky coverage, but they have the same channel bandwidth and we have been consistent in the DM range that was searched.
Each of these four projects have successfully detected at least one FRB, as listed in Table~\ref{table:collections}. 

Using the FRB event rate predicted by~\citet{Champion16}, approximately 12 FRBs would have been detected in a HTRU-style survey~\citep{Keith10} assuming the same integration time as our observations. However, the more modern surveys use signal processors with channel widths of only 0.39\,MHz compared with 3\,MHz for the data processed in this paper.  In Figure~\ref{figure:Parkes_FRB} we present the start date (using a vertical line) of the HTRU survey, which made use of the narrower frequency channels.  The use of wider channel widths implies that the observations are less sensitive to high-DM FRB events.  For instance, the DM-smearing across a 3\,MHz channel is $\sim$9\,ms at 1.4\,GHz for an FRB event with a DM of 1000\;cm$^{-3}\,$pc. Considering the DMs and observed pulse widths for the 28 FRBs that have been discovered using the Parkes telescope, we find that $\sim$14 would have been missed (or difficult to detect because of significantly lower S/N values) in the earlier data.  The event rate of detectable FRBs in the earlier data is therefore significantly lower than in the more recent data.  After accounting for our wider channel bandwidths we expect $\sim$5 FRB events in the earlier data (compared with five that we have detected). Note that our one-bit sampled data sets and the sampling time (for some of the data sets) of 1\,ms can also affect the detection rate of our search. Our detections are therefore consistent with the FRB event prediction. 

Using the definition in \citet{Bhandari18} of three regions delineated in Galactic latitude of $|$b$|$ $\leq$ 19.5$^{\circ}$, 19.5$^{\circ}$  $<$  $|$b$|$ $\le$  42$^{\circ}$ and 42$^{\circ}$  $<$  $|$b$|$, we list the on-sky integration time in these three latitude ranges for our search in Table~\ref{table:obs_len_ALT1}. Four of the five FRBs reported in our data sets occurred in the intermediate-latitude region that contained 3,582\,hr observations from Jan. 2001 to Aug. 2001. The remaining FRB (FRB~010621) occurred in the low-latitude region that contained 5,016\,hr observations in the same time range. The single detection of an FRB in this region can be explained by the effect of Galactic diffractive interstellar scintillation~\citep{Macquart15}.

Is it reasonable to detect five FRB events within six months, but no event in the previous 3.5 years? 
Our search contained 28,690\,hr of observations prior to Jan. 2001 and 9,500\,hr of observations from Jan. 2001 to Aug. 2001 (see Table~\ref{table:obs_len_ALT1} for more details).
Assuming that the occurrence of FRBs has a  Poisson distribution,  the probability of observing $k$ events in a given interval is:
\begin{equation}
P(X=k) = e^{-\lambda} \frac{\lambda^{k}}{k!},
\label{equ:poisson}
\end{equation}
where $\lambda$ is the mean number of events per interval. Therefore the probability of detecting no event in our 28,690\,hr observations prior to Jan. 2001 ($\lambda = 3.756$) is $\sim$ 0.0234, corresponding to a confidence level of 0.976, i.e., close to 2\,$\sigma$ level of significance that the FRBs are unevenly distributed in time~\citep{Gehrels86}.
%close to 2\,$\sigma$ level of significance.

\begin{table*}
\caption{The on-sky integration time, in three latitude ranges, for observations before Jan 2001 and from Jan. to Aug. 2001. Four of the five FRBs reported in our data set occurred in the intermediate-latitude region from Jan. 2001 to Aug. 2001 and the remaining FRB (FRB~010621) occurred in the low-latitude region in the same time range.}
\begin{center}
\begin{tabular}{lcccr}
\hline
\hline
\multicolumn{1}{c}{Observational} & \multicolumn{3}{c}{Galactic latitude}                                                                                                                                                                      & \multicolumn{1}{c}{Total} \\\cline{2-4}%\\
\multicolumn{1}{c}{Period}        & \multicolumn{1}{c}{$0^{\circ} \leq |\textrm{b}| \leq 19.5^{\circ}$}  &  \multicolumn{1}{c}{$19.5^{\circ}  <  |\textrm{b}| \le  42^{\circ}$}  &  \multicolumn{1}{c}{$42^{\circ}  <  |\textrm{b}| \le  90^{\circ}$}  & \multicolumn{1}{c}{(h)}   \\
%\multicolumn{1}{c}{Period}        & \multicolumn{1}{c}{$[0^{\circ},19.5^{\circ}]$}                       &  \multicolumn{1}{c}{$(19.5^{\circ}$,42$^{\circ}]$}                    &  \multicolumn{1}{c}{$(42^{\circ},90^{\circ}]$}                      & \multicolumn{1}{c}{(h)}   \\
\hline
Prior to Jan. 2001      & \makebox[1cm][r]{24,922} & \makebox[1cm][r]{2,217} & \makebox[0.9cm][r]{1,551} & 28,690  \\
Jan. 2001 --- Aug. 2001 & \makebox[1cm][r]{ 5,016} & \makebox[1cm][r]{3,582} & \makebox[0.9cm][r]{  902} &  9,500  \\
Complete                & \makebox[1cm][r]{29,938} & \makebox[1cm][r]{5,799} & \makebox[0.9cm][r]{2,453} & 38,190  \\
\hline
\end{tabular}
\end{center}
\label{table:obs_len_ALT1}
\end{table*}

\subsection{Using the database to produce a training data set}
\label{sec:db_use_training}
%usage of the database: pulsar study; software comparing; ML training; make the raw data smaller; no need for downlaod and process; chance to find new event;
Our database contains 568,736,756 single pulse candidates, too many to view by eye.  It is therefore likely that large-scale searches of the database will be carried out using machine learning or matched-filtering algorithms.  Many such algorithms rely on data sets that can be used to train or optimise the algorithms and the labelled candidates in our database provides such a data set.

Known pulsars are labelled directly within the database.  Currently we have labelled 385 pulsars for which single pulses are detectable. These pulsars have a range of properties with pulse period from 0.016 to 1.98\,s and dispersion measures between 2.64 and 1172.0\,cm$^{-3}\,$pc. Descriptions of how to identify the relevant data files and segments are given in the Appendix~\ref{ap_usage}.  

In the Appendix~\ref{ap_usage} we also show how a user can extract the five FRB events and the single pulse candidates that have been described in this paper. We have also identified 10,614 segments of data that exhibit a range of RFI signatures (including 22 perytons) and 100 segments in which weak (S/N $\sim$ 7) burst events were recorded. Finally, we list 14,017 files in which no pulsed candidates were identified and therefore represent ``clean'' data.

\section{Conclusion}
\label{sec:conclu}
%database usage discussion 
As we have re-processed all the data sets with a self-consistent search, the gap of the FRBs detection is unlikely caused by the events not being identified during previous searches.  The paucity of FRB in the early data sets can be explained by the large DM-smearing and the uneven distribution on time is only a fluctuation of $\sim$2 sigma.

Pulsars, RRATs and FRBs continue to be discovered in both new and archival data sets.  In the near future the data rates from the next generation of telescopes will be so high that the raw data files will unlikely be able to be preserved.  Real-time algorithms will therefore process the data and top-ranked candidates presented to the astronomers.  It is likely that a database of the pulsed candidates will be produced, and we have explored such a database in this paper.  We note that our database is currently 41\,GiB, which should be compared with 5.7\,TiB for the raw data. 

We have demonstrated how such databases can be used to discover new sources. Only $\sim$ 199\,hr of the total 38,190\,hr (i.e., $\sim$ 0.52\%) of the data sets contain candidates. The candidates are dominated by single pulses from known pulsars or RFI signals.  We have chosen to also keep the candidate caused by RFI events in our database as we wish to carry out a long-term study of impulsive RFI at the Parkes observatory, but methods to identify the RFI (such as multibeam information; or using machine learning algorithms trained to detect RFI) could significantly reduce the size of the database.

%next version & final version
This is the first version of the database of the single pulses from Parkes telescope as we only searched a part of the archival observations. More data sets of the data archive are being processed and will be included in next version. The database is publicly available.  We hope it will be a useful resource for both the pulsar and FRB communities.

\section*{Acknowledgments}
The Parkes radio telescope is part of the Australia Telescope National Facility which is funded by the Australian Government for operation as a National Facility managed by CSIRO. This paper includes archived data obtained through the CSIRO Data Access Portal (http://data.csiro.au). This project was supported by resources and expertise provided by CSIRO IMT Scientific Computing. This work was supported by a China Scholarship Council (CSC) Joint PhD Training Program grant, the National Natural Science Foundation of China (Grant No. 11725314) and the Guizhou Provincial Science and Technology Foundation (Grant No. [2020]1Y019). Parts of this research were supported by the Australian Research Council Centre of Excellence for All Sky Astrophysics in 3 Dimensions (ASTRO 3D), through project number CE170100013.

\appendix

\section{Downloading and using the database}
\label{sec:usage}

\subsection{Downloading}
The database is available from:
\begin{quote}
\url{https://data.csiro.au/dap/landingpage?pid=csiro:42640}
\end{quote}

When connecting to the DAP, because of the size of the file, you will request access via an email address.  A temporary password will be sent to your email to access the file directly.  You need to request access via WebDAV.

In the email you will be able to click through the final location.  You can click through to the file in most contemporary web browsers and 64-bit operating systems.  Again, because of the size of the file (and the idiosyncrasies of different web browsers and operating systems) it may be easier to download with a scripting tool like \texttt{wget}.  For example:
\begin{quote}
\texttt{wget -m -np -nH --user=USER --password=PASSWORD
$\backslash$\newline
https://webdav-bm.data.csiro.au/dap\_prd\_000042640v001/}
\end{quote}
replacing USER and PASSWORD with the User and Password fields in the email.

\subsection{Statistics}
\label{sec:stat}

Currently:
\begin{itemize}
    \item it is distributed as a tar-ball (\texttt{database\_file.tar.gz});
    \item the archive is $\sim$~9.3\,GiB;
    \item when extracted the SQLite database:
    \begin{itemize}
        \item requires $\sim$~41\,GiB for the single database file (\texttt{singlePulseDAP.db});
        \item has two (2) additional support directories:
        \begin{itemize}
            \item \texttt{db\_fileSeg} containing 6,944 file segments, requiring 573\,MiB of disk space;
            \item \texttt{db\_image} containing 6,944 corresponding PNG files, requiring 235\,MiB of disk space.
        \end{itemize}
    \end{itemize}
\end{itemize}

\subsection{Usage} % or Tools
\label{ap_usage}
The database can be accessed using \emph{\sc SQLite} directly.  We also provide six (6) software tools to manage the database:
\begin{itemize}
    \item \texttt{plot\_TimeDM} to plot the DM$-$time plane of one file/observation;
    \item \texttt{pfits\_read\_1bitExtraction} to plot the frequency$-$time plane of one file segment using the extracted files;
    \item \texttt{cone\_check\_PSR.py} to provide pulsars close to the beam pointing position;
    \item \texttt{cone\_check\_SPC.py} to do a cone search for the candidates in the database;
    \item \texttt{searchFRB.py} will output the filenames, or the fileIDs and fileSegmentIDs of who containing FRB-like phenomena;
    \item \texttt{get\_training\_data.py} can output the information of different kinds of data set:
    \begin{itemize}
        \item FRB;
        \item perytons;
        \item pulsar (DM and flux ranges can be specified).
    \end{itemize}
\end{itemize}    

To compile and execute the code:
\begin{enumerate}
    \item Download the source code from the git repository:
    \begin{quote}
        \url{https://bitbucket.csiro.au/scm/spdb/spc-usertools.git}
    \end{quote}
    with a web-browser or using an appropriate git tool.  On the command line this might look like:
%    \begin{quote}
%        \begin{center}
%            \texttt{git clone $\backslash$\newline https://bitbucket.csiro.au/scm/spdb/spc-usertools.git}
%        \end{center}
%    \end{quote}
    \begin{quote}
        \begin{center}
            \texttt{git clone https://bitbucket.csiro.au/scm/spdb/spc-usertools.git}
        \end{center}
    \end{quote}

    \item run \texttt{make} to compile the C code, the dependencies are: \emph{\sc sqlite3}, \emph{\sc cfitsio} and \emph{\sc pgplot};
    \item Python code should run within a Python~3 environment;
    \item use the \texttt{-h} command option for the supplied programs to print the help page;
    \item for instance, to get the training data for pulsars with DM from 100 to 110\,cm$^{-3}\,$pc and flux at 1400\,MHz from 1.0 to 2.0\,mJy:
%        \begin{quote}
%            \begin{center}
%                \texttt{get\_training\_data -t pulsar $\backslash$\newline
%                -dml 100 -dmh 110 -fl 1 -fh 2}
%            \end{center}
%        \end{quote}
        \begin{quote}
            \begin{center}
                \texttt{get\_training\_data -t pulsar 
                -dml 100 -dmh 110 -fl 1 -fh 2}
            \end{center}
        \end{quote}
    generating the collection and filename information, for example:\\
\begin{tabular}{l}
\texttt{Jname, filename, collection}\\
\texttt{J1059-5742 PM0005\_00881.sf 1997AUGT-P268}\\
\texttt{J1059-5742 PM0005\_012D1.sf 1997AUGT-P268}\\
\texttt{J1059-5742 PM0007\_02251.sf 1997AUGT-P268}\\
\texttt{J1059-5742 PM0016\_02081.sf 1997AUGT-P268}
\end{tabular}   
\end{enumerate}

\section{DOIs of the data collections used in this study}
\label{raw_data_dois}
\begin{center}
\begin{tabular}{ll}
\hline
\hline
Collection Name &  DOI  \\
\hline
P268$-$1997AUGT  &     doi:10.4225/08/583746ac2c4de  \\
P268$-$1998JANT  &     doi:10.4225/08/5850b59a36ecb  \\
P268$-$1998MAYT  &     doi:10.4225/08/5850b61f44170  \\
P268$-$1998SEPT  &     doi:10.4225/08/587b1bab111be  \\
P309$-$1998SEPT  &     doi:10.4225/08/577DD57C0F90F  \\
P268$-$1999JANT  &     doi:10.4225/08/587b1bd633e6b  \\
P309$-$1999JANT  &     doi:10.4225/08/577F5A8B0CB63  \\
P268$-$1999MAYT  &     doi:10.4225/08/587ddb7d67a88  \\
P309$-$1999MAYT  &     doi:10.4225/08/578836214A3A5  \\
P268$-$1999SEPT  &     doi:10.4225/08/5884ff9cd4cad  \\
P268$-$2000JANT  &     doi:10.4225/08/5884ffd19c975  \\
P268$-$2000MAYT  &     doi:10.4225/08/5885000013669  \\
P269$-$2000MAYT  &     doi:10.4225/08/5819187e513a7  \\
P268$-$2000OCTT  &     doi:10.4225/08/58a0ff51a47ab  \\
P269$-$2000OCTT  &     doi:10.4225/08/5819364264904  \\
P268$-$2001JANT  &     doi:10.4225/08/58a3aa8588ead  \\
P269$-$2001JANT  &     doi:10.4225/08/5819628e4fed9  \\
P360$-$2001JANT  &     doi:10.4225/08/5805cc2801cf7  \\
P268$-$2001MAYT  &     doi:10.4225/08/58a3aab36c1a2  \\
P360$-$2001MAYT  &     doi:10.4225/08/5807812df0644  \\
P366$-$2001MAYT  &     doi:10.4225/08/598c2d9103f0c  \\
\hline
\hline
\end{tabular} 
\end{center}


\begin{thebibliography}{99}

\bibitem[Bannister et al.(2019)]{Bannister19} Bannister, K.~W., Deller, A.~T., Phillips, C., et al.\ 2019, Science, 365, 565

\bibitem[Bates et al.(2011)]{Bates11} Bates, S.~D., Johnston, S., Lorimer, D.~R., et al.\ 2011, \mnras, 411, 1575

\bibitem[Bhandari et al.(2018)]{Bhandari18} Bhandari, S., Keane, E.~F., Barr, E.~D., et al.\ 2018, \mnras, 475, 1427

\bibitem[Burgay et al.(2006)]{Burgay06} Burgay, M., Joshi, B.~C., D'Amico, N., et al.\ 2006, \mnras, 368, 283

\bibitem[Burke-Spolaor \& Bannister(2014)]{Burke-Spolaor14} Burke-Spolaor, S., \& Bannister, K.~W.\ 2014, \apj, 792, 19

\bibitem[Champion et al.(2016)]{Champion16} Champion, D.~J., Petroff, E., Kramer, M., et al.\ 2016, \mnras, 460, L30

\bibitem[Cordes \& McLaughlin(2003)]{Cordes03} Cordes, J.~M., \& McLaughlin, M.~A.\ 2003, \apj, 596, 1142

\bibitem[Crawford et al.(2001)]{Crawford01} Crawford, F., Kaspi, V.~M., Manchester, R.~N., et al.\ 2001, \apj, 553, 367

\bibitem[Cui et al.(2017)]{Cui17} Cui, B.-Y., Boyles, J., McLaughlin, M.~A., et al.\ 2017, \apj, 840, 5

\bibitem[Edwards et al.(2001)]{Edwards01} Edwards, R.~T., Bailes, M., van Straten, W., et al.\ 2001, \mnras, 326, 358

\bibitem[Gehrels(1986)]{Gehrels86} Gehrels, N.\ 1986, \apj, 303, 336

\bibitem[Hobbs et al.(2004)]{Hobbs04} Hobbs, G., Faulkner, A., Stairs, I.~H., et al.\ 2004, \mnras, 352, 1439

\bibitem[Hobbs et al.(2011)]{Hobbs11} Hobbs, G., Miller, D., Manchester, R.~N., et al.\ 2011, \pasa, 28, 202

\bibitem[Hobbs et al.(2020)]{Hobbs19} Hobbs, G., Manchester, R.~N., Dunning, A., et al.\ 2020, \pasa, in press (arXiv:1911.00656)

\bibitem[Jacoby et al.(2009)]{Jacoby09} Jacoby, B.~A., Bailes, M., Ord, S.~M., et al.\ 2009, \apj, 699, 2009

\bibitem[Jankowski et al.(2018)]{Jankowski18} Jankowski, F., van Straten, W., Keane, E.~F., et al.\ 2018, \mnras, 473, 4436

\bibitem[Jiang et al.(2019)]{Jiang19} Jiang, P., Yue, Y., Gan, H., et al.\ 2019, Science China Physics, Mechanics, and Astronomy, 62, 959502

\bibitem[Jonas \& MeerKAT Team(2016)]{Jonas16} Jonas, J., \& MeerKAT Team\ 2016, Proc. Sci., The MeerKAT Radio Telescope. SISSA, Trieste, PoS(MeerKAT2016)001

\bibitem[Katz(2016)]{Katz16} Katz, J.~I.\ 2016, \apj, 818, 19

\bibitem[Keane et al.(2012)]{Keane12} Keane, E.~F., Stappers, B.~W., Kramer, M., et al.\ 2012, \mnras, 425, L71

\bibitem[Keane et al.(2019)]{Keane19} Keane, E.~F., Lorimer, D.~R., \& Crawford, F.\ 2019, Research Notes of the American Astronomical Society, 3, 41

\bibitem[Keith et al.(2010)]{Keith10} Keith, M.~J., Jameson, A., van Straten, W., et al.\ 2010, \mnras, 409, 619

\bibitem[Kramer et al.(2003)]{Kramer03} Kramer, M., Bell, J.~F., Manchester, R.~N., et al.\ 2003, \mnras, 342, 1299

\bibitem[Lorimer et al.(2006)]{Lorimer06} Lorimer, D.~R., Faulkner, A.~J., Lyne, A.~G., et al.\ 2006, \mnras, 372, 777

\bibitem[Lorimer et al.(2007)]{Lorimer07} Lorimer, D.~R., Bailes, M., McLaughlin, M.~A., et al.\ 2007, Science, 318, 777

\bibitem[Macquart \& Johnston(2015)]{Macquart15} Macquart, J.-P., \& Johnston, S.\ 2015, \mnras, 451, 3278

\bibitem[Manchester et al.(2001)]{Manchester01} Manchester, R.~N., Lyne, A.~G., Camilo, F., et al.\ 2001, \mnras, 328, 17

\bibitem[Manchester et al.(2005)]{Manchester05} Manchester, R.~N., Hobbs, G.~B., Teoh, A., et al.\ 2005, \aj, 129, 1993

\bibitem[Manchester et al.(2006)]{Manchester06} Manchester, R.~N., Fan, G., Lyne, A.~G., et al.\ 2006, \apj, 649, 235

\bibitem[McLaughlin et al.(2006)]{McLaughlin06} McLaughlin, M.~A., Lyne, A.~G., Lorimer, D.~R., et al.\ 2006, \nat, 439, 817

\bibitem[Morris et al.(2002)]{Morris02} Morris, D.~J., Hobbs, G., Lyne, A.~G., et al.\ 2002, \mnras, 335, 275

\bibitem[Os{\l}owski et al.(2019)]{Oslowski19} Os{\l}owski, S., Shannon, R.~M., Ravi, V., et al.\ 2019, \mnras, 488, 868

\bibitem[Pan et al.(2016)]{Pan16} Pan, Z., Hobbs, G., Li, D., et al.\ 2016, \mnras, 459, L26

\bibitem[Petroff et al.(2015)]{Petroff15} Petroff, E., Keane, E.~F., Barr, E.~D., et al.\ 2015, \mnras, 451, 3933

\bibitem[Petroff et al.(2016)]{Petroff16} Petroff, E., Barr, E.~D., Jameson, A., et al.\ 2016, \pasa, 33, e045

\bibitem[Petroff et al.(2019)]{Petroff19} Petroff, E., Oostrum, L.~C., Stappers, B.~W., et al.\ 2019, \mnras, 482, 3109

\bibitem[Ransom(2001)]{Ransom01} Ransom, S.~M.\ 2001, Ph.D. Thesis, Harvard University 

\bibitem[Shannon et al.(2018)]{Shannon18} Shannon, R.~M., Macquart, J.-P., Bannister, K.~W., et al.\ 2018, \nat, 562, 386

\bibitem[Spitler et al.(2016)]{Spitler16} Spitler, L.~G., Scholz, P., Hessels, J.~W.~T., et al.\ 2016, \nat, 531, 202

\bibitem[Staveley-Smith et al.(1996)]{Staveley-Smith96} Staveley-Smith, L., Wilson, W.~E., Bird, T.~S., et al.\ 1996, \pasa, 13, 243

\bibitem[The CHIME/FRB Collaboration et al.(2019a)]{CHIME19a} The CHIME/FRB Collaboration, Amiri, M., Bandura, K., et al.\ 2019a, \nat, 566, 230

\bibitem[The CHIME/FRB Collaboration et al.(2019b)]{CHIME19b} The CHIME/FRB Collaboration, Andersen, B.~C., Bandura, K., et al.\ 2019b, \apjl, 885, L24

\bibitem[The Planck Collaboration et al.(2014)]{Planck14} The Planck Collaboration, Ade, P.~A.~R., Aghanim, N., et al.\ 2014, \aap, 571, A16

\bibitem[Thornton et al.(2013)]{Thornton13} Thornton, D., Stappers, B., Bailes, M., et al.\ 2013, Science, 341, 53

\bibitem[Yao et al.(2017)]{Yao17} Yao, J.~M., Manchester, R.~N., \& Wang, N.\ 2017, \apj, 835, 29

\bibitem[Zhang et al.(2018)]{Zhang18} Zhang, S.-B., Dai, S., Hobbs, G., et al.\ 2018, \mnras, 479, 1836

\bibitem[Zhang et al.(2019)]{Zhang19} Zhang, S.-B., Hobbs, G., Dai, S., et al.\ 2019, \mnras, 484, L147

\end{thebibliography}
\end{document}